\documentclass[11pt]{article}

\usepackage{amsmath}
\usepackage{amssymb}
\usepackage{amscd}

\usepackage[english]{babel}

\usepackage[raiselinks=false,plainpages=false,pdfpagelabels,naturalnames=true,colorlinks=true,citecolor=blue,urlcolor=blue,linkcolor=blue,bookmarksopen=true,pagebackref,hyperindex]{hyperref}

\allowdisplaybreaks[1]

\usepackage{latexsym}
\usepackage[emu]{cmll}
\usepackage{prooftree}

\def\Section#1{\section{#1}}
\def\Subsection#1{\subsection{#1}}

\def\Hide#1{\relax}

\DeclareSymbolFont{AMSb}{U}{msb}{m}{n}
\DeclareSymbolFontAlphabet{\mathbb}{AMSb}
\DeclareSymbolFont{symbolsC}{U}{txsyc}{m}{n}
\DeclareMathSymbol{\rJoin}{\mathrel}{symbolsC}{89}

\newcommand{\CK}{\vee}
\newcommand{\oCK}{\downarrow}
\newcommand{\CM}{\Rightarrow}
\newcommand{\CC}{\wedge}

\newcommand{\CKb}[2]{#1\vee{#2}}
\newcommand{\oCKb}[2]{#1\downarrow{#2}}
\newcommand{\CMb}[2]{#1\Rightarrow{#2}}
\newcommand{\CCb}[2]{#1\wedge{#2}}

\def\VA{\mathbf{A}}

\def\VT{\mathbf{T}}

\def\Func#1{{\sf{#1}}}

\def\iAnd{\otimes} 

\newtheorem{Theorem}{Theorem}[subsection]
\newtheorem{Lemma}[Theorem]{Lemma}
\newtheorem{Corollary}[Theorem]{Corollary}

\newtheorem{Definition}{Definition}[subsection]
\newtheorem{Remark}{Remark}[subsection]

\def\Proof{\par \noindent{\bf Proof: }}
\def\Done{\hfill\rule{0.5em}{0.5em}}
\def\Var{\Func{Var}}

\def\lL{{\cal L}}

\def\ASM{[\Func{ASM}]}

\def\CE{[{\iAnd}\Func{E}]}
\def\CI{[{\iAnd}\Func{I}]}

\def\CONR{[\Func{CON}_r]}

\def\CON{[\Func{CON}]}

\def\CWC{[\Func{CWC}]}
\def\MWC{[\Func{MWC}]}
\def\AWC{[\Func{AWC}]}
\def\DNE{[\Func{DNE}]}

\def\EFQ{[\Func{EFQ}]}
\def\EFQ{[\Func{EFQ}]}

\def\LE{[{\Lolly}\Func{E}]}
\def\LI{[{\Lolly}\Func{I}]}

\def\Lnot{{{}^{\perp}}}
\def\Lolly{\multimap}

\def\WC{[\Func{wCON}]}
\def\WK{[\Func{WK}]}

\def\axWk{(\Func{Wk})}
\def\axComp{(\Func{Comp})}
\def\axComm{(\Func{Comm})}
\def\axUncurry{(\Func{Uncurry})}
\def\axCurry{(\Func{Curry})}
\def\axEFQ{(\Func{EFQ})}
\def\axDNE{(\Func{DNE})}
\def\axCWC{(\Func{CWC})}
\def\axCon{(\Func{Con})}

\def\Logic#1#2{\mbox{{\bf #1}}_{\mbox{\bf #2}}}

\def\ALi{\Logic{AL}{i}}
\def\ALm{\Logic{AL}{m}}
\def\BL{\Logic{BL}{}}
\def\ALc{\Logic{AL}{c}}
\def\Luk{\Logic{{\L}}{\relax}}
\def\LLi{\Logic{{\L}L}{i}}
\def\LLm{\Logic{{\L}L}{m}}
\def\LLc{\Logic{{\L}L}{c}}

\def\IL{\Logic{IL}{}}
\def\ML{\Logic{ML}{}}

\def\rImp{\mathop{\rightarrow}}

\newcommand{\by}[1]{\tag*{(#1)}}

\def\Hilb#1{#1^{{\rm h}}}

\newcommand{\glTrans}[1]{{#1}^{\sf Gli}}

\title{On Affine Logic and {\L}ukasiewicz Logic}

\author{Rob Arthan \& Paulo Oliva}

\begin{document}
\maketitle

\begin{abstract}
The multi-valued logic of {\L}ukasiewicz is a substructural logic that has been
widely studied and has many interesting properties. It is classical, in the
sense that it admits the axiom schema of double negation elimination, $\DNE$. However, our
understanding of {\L}ukasiewicz logic can be improved by separating its
classical and intuitionistic aspects.  The intuitionistic aspect of
{\L}ukasiewicz logic is captured in an axiom schema, $\CWC$, which asserts the
commutativity of a weak form of conjunction. We explain how this axiom is equivalent to a
restricted form of contraction.  We then show how {\L}ukasiewicz Logic can be viewed
both as an extension of classical affine logic with $\CWC$, or as an extension
of what we call \emph{intuitionistic} {\L}ukasiewicz logic with $\DNE$,
intuitionistic {\L}ukasiewicz logic being the extension of intuitionistic
affine logic by the schema $\CWC$.  At first glance, intuitionistic {\L}ukasiewicz
logic seems to be a very weak fragment of intuitionistic logic. We show that $\CWC$ is a
surprisingly powerful (restricted) form of contraction, implying
for instance all the intuitionistically valid De Morgan's laws. However the
proofs can be very intricate. These results are presented using derived
connectives to clarify and motivate the proofs. Applications include:
a simpler proof of the Ferreirim-Veroff-Spinks theorem, proof that idempotent elements of a hoop form a sub-hoop,
proof that double negation is a hoop homomorphism, and proofs for the above mentioned De Morgan dualities.
We conclude by showing that our homomorphism result on the double-negation mapping in particular implies
that all negative translations of classical into intuitionistic {\L}ukasiewicz coincide
(as they do in full intuitionistic logic). 
This is in contrast with affine logic for which we show, by appeal to results on semantics
proved in a companion paper, that both the Gentzen and the Glivenko
translations fail. 

\end{abstract}

\Section{Introduction}
\label{sec:introduction}

This paper studies two fragments of {\L}ukasiewicz logic $\LLc$, which we have called \emph{minimal {\L}ukasiewicz logic} $\LLm$ and \emph{intuitionistic {\L}ukasiewicz logic} $\LLi$. Just as {\L}ukasiewicz logic \cite{Hajek98} is a subsystem of classical (Boolean) logic, minimal and intuitionistic {\L}ukasiewicz logic are subsystems of minimal and intuitionistic logic, respectively \cite{Troelstra(96)}. Our approach, however, is the study these three systems as extensions of \emph{minimal affine logic}, i.e. the fragment of linear logic \cite{Girard(87B)} containing implication ($\Lolly$) and the multiplicative conjunction ($\iAnd$) together with the weakening rule but without the constant for falsehood, and hence without negation. A similar sequent calculus for {\L}ukasiewicz logic based on affine logic has been proposed in \cite{Ciabattoni:1997}. The main differences are that we work on the implication-conjunction fragment of affine logic, and take \emph{minimal} affine logic as the starting point. As it will be clear, we also focus on the minimal and intuitionistic fragments of {\L}ukasiewicz logic, and aim to prove theorems \emph{of} these system, rather than theorems \emph{about} the system (meta-theorems).

As shown in Figure \ref{fig:logics}, we shall view \emph{minimal {\L}ukasiewicz logic} $\LLm$ as sitting in between minimal affine logic $\ALm$ and (the usual) minimal logic $\ML$. Starting from one of the ``minimal" fragments, an intuitionistic variant is obtained by adjoining the constant for falsehood ($1$ in our case) and the principle \emph{ex falso quodlibet}
\[ 1 \vdash A \]
The adjoining of ex falso quodlibet takes us from the lowest row in Figure \ref{fig:logics} to the middle row where again intuitionistic {\L}ukasiewicz logic $\LLi$ can be placed between {intuitionistic affine logic $\ALi$ and the usual (implication-conjunction fragment of) intuitionistic logic $\IL$. In the intuitionistic systems we consider, negation will be defined via implication and falsehood as $A\Lnot = A \Lolly 1$. 

From the intuitionistic systems we can obtain ``classical" counterparts by adding the law of \emph{double negation elimination}
\[ A\Lnot\Lnot \vdash A \]
again moving us one level up in the diagram above. In order to move horizontally on the diagram from the left-most column (affine system) to the right-most column (minimal $\ML$, intuitionistic $\IL$ and Boolean logic $\BL$) one adds the \emph{contraction axiom}
\[ A \vdash A \iAnd A \]
The {\L}ukasiewicz logical systems sit in between affine systems, where no contraction is permitted, and the full systems which contain the contraction axiom for all formulas.

\begin{figure}[t]
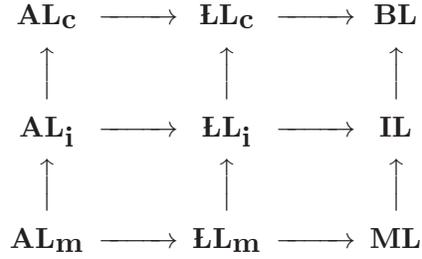

\[
\begin{CD}
\ALc  @>>> \LLc  @>>> \BL \\
@AAA       @AAA       @AAA \\
\ALi  @>>> \LLi  @>>> \IL \\
@AAA       @AAA       @AAA \\
\ALm  @>>> \LLm  @>>> \ML \\
\end{CD}
\]
\caption{Relationships between the Logics}
\label{fig:logics}
\end{figure}

One of our motivations for the present work was to identify precisely what amount of contraction needs to be added to affine logic, in order to obtain {\L}ukasiewicz logic. As the presence of classical logic can often obscure the role of contraction, we were naturally led in our investigation to move to systems where no classical logic (double negation elimination) and even no falsehood constant is assumed, leading us to this minimal logic variant of {\L}ukasiewicz logic.

What we have discovered is that the three {\L}ukasiewicz systems $\LLm, \LLi$ and $\LLc$, can be obtained from their affine counterparts (namely $\ALm, \ALi$ and $\ALc$) by adjoining the axiom of \emph{commutativity of weak conjunction} 
\[ A \iAnd (A \Lolly B) \vdash B \iAnd (B \Lolly A) \]
which is a simple consequence of contraction, but is strictly weaker than it. This is the well-known \emph{axiom of divisibility}, so called because it says intuitively that when $A$ is greater than $B$ (i.e. $A \Lolly B$) then $A$ can divided into two components namely $B$ and $B \Lolly A$.

The reason we call $A \iAnd (A \Lolly B)$ a weak form of conjunction can be explained as follows: Note that $A \iAnd (A \Lolly B)$ implies both $A$ and $B$, but without contraction (so that $A$ can be used twice), we do not have in general $A \iAnd (A \Lolly B) \vdash A \iAnd B$. On the other hand, due to the presence of weakening in the affine systems, we always have $A \iAnd B \vdash A \iAnd (A \Lolly B)$. Hence, $A \iAnd (A \Lolly B)$ is strictly weaker than the usual multiplicative conjunction $A \iAnd B$. The axiom above states that this conjunction is commutative. 

This permitted us to separate the ``amount of classical logic" available in $\LLc$ from the weak form of contraction that is present there, allowing us to see {\L}ukasiewicz logic $\LLc$ as the combination of classical affine logic. $\ALc$ with minimal {\L}ukasiewicz logic $\LLm$. The first system contains full double negation elimination $\DNE$ but no contraction, whereas the second does not valid $\DNE$ in general but contains a certain amount of contraction as expressed by the commutativity of weak conjunction.

The intuitionistic component of $\LLc$, which we call intuitionistic {\L}ukasiewicz Logic $\LLi$ has been studied before, under different names. For instance, Blok and Ferreirim \cite{blok-ferreirim00} refer to it as ${\sf S}_{\cal HO}$.

We express here our gratitude to the late Bill McCune for the development of
Prover9 and Mace4, which we have extensively used in both trying to find
derivations or counter-examples to our various conjectures. Most of the
derivations found in this paper were initially found by Prover9. The authors'
contribution was to propose conjectures to Prover9, to study the
machine-oriented proofs it found and to present the proofs in a
human-intelligible form by breaking them down into structurally interesting
lemmas. This was an iterative process since often Prover9 was able to find
simpler proofs of a lemma when presented with it as a conjecture in isolation.
In cases when Prover9 was unable to find a proof, Mace4 was often able to find
a counter-example: a finite model of the logic in question in which the
conjecture can be seen to fail. See our companion paper~\cite{arthan-oliva14b}
for examples of models found by Mace4.

In fact, this second component of our work, namely extracting ``meaning" from 
the machine-oriented proofs found by Prover9, led to the identification of four important \emph{derivable} connectives that showed up again and again in the proofs. These are,
\[
\begin{array}{lcll}
	\CCb{A}{B} & \equiv & A \iAnd (A \Lolly B) & \mbox{(weak conjunction)} \\[2mm]
	\CKb{A}{B} & \equiv & (B \Lolly A) \Lolly A & \mbox{(strong disjunction)} \\[2mm]
	\CMb{A}{B} & \equiv & A \Lolly A \iAnd B & \mbox{(strong implication)} \\[2mm]
	\oCKb{A}{B} & \equiv & A\Lnot \iAnd (B \Lolly A) \quad \quad & \mbox{(NOR binary connective)}
\end{array}
\]
It is easy to check that classically these are indeed equivalent to $A$ and $B$, $A$ or $B$, $A$ implies $B$, and neither $A$ nor $B$, respectively. In Sections \ref{sec:weak-contraction} and \ref{sec-double-negation} we prove importants results in $\LLm$ and $\LLi$, respectively, most of these are properties relating the various derived connectives or relating derived connectives with the primitive connectives, $\iAnd$ and $\Lolly$. 

We give several applications of the our results, including: (1) A simpler and more abstract version of a proof, originally found by Veroff and Spinks \cite{Veroff01}, of one of Ferreirim's theorem \cite{Ferreirim92} (cf. Section \ref{sec:f-v-s-theorem}), (2) a proof that the idempotent elements of a hoop form a sub-hoop (cf. Section \ref{sec:idem}), (3) homomorphism properties for the double negation operator in intuitionistic {\L}ukasiewicz logic (cf. Sections \ref{homo-dn-lolly} and \ref{homo-dn-tensor}), and (4) a collection of De Morgan properties for intuitionistic {\L}ukasiewicz logic (cf. Section \ref{sec-de-morgan}). 

Finally, in Section \ref{sec:embedding} we make use of the homomorphim properties of the double negation to show that both the Gentzen and the Glivenko translations, while they fail for affine logic, surprisingly still work as a double negation translation of classical {\L}ukasiewicz logic into intuitionistic {\L}ukasiewicz logic.

The reader with model-theoretic inclinations is invited to consult our companion paper \cite{arthan-oliva14b} in which we investigate algebraic semantics for the logics studied here. This leads to a novel indirect method for demonstrating that a formula is provable in {\L}ukasiewicz Logic, a method which we have used extensively in parallel with attempts to find explicit proofs and counter-examples with Prover9 and Mace4.

\begin{figure}[t]
\begin{center}
\begin{tabular}{|cc|}\hline
& \\
%
%
\begin{prooftree}
\Gamma, A \vdash B
\justifies
\Gamma \vdash A \Lolly B
\using{\LI}
\end{prooftree} &
%
%
\begin{prooftree}
\Gamma \vdash A \quad \Delta \vdash A \Lolly B
\justifies
\Gamma, \Delta \vdash B
\using{\LE}
\end{prooftree} \\\ &  \\
%
%
\; \begin{prooftree}
\Gamma \vdash A \quad \Delta \vdash B
\justifies
\Gamma, \Delta \vdash A \iAnd B
\using{\CI}
\end{prooftree}
&
%
%
\begin{prooftree}
\Gamma \vdash A \iAnd B \quad \Delta, A, B \vdash C
\justifies
\Gamma, \Delta \vdash C
\using{\CE}
\end{prooftree} \\[6mm]
\hline
\end{tabular}
\caption{Sequent Calculus Rules}
\label{fig:sequent-rules}
\end{center}
\end{figure}

\Section{Definitions of the Logics}
\label{sec:logics-and-algebras}

We work in a language, $\cal L$, built from a countable set of propositional variables
$\Var = \{V_1, V_2, \ldots\}$, the constant $1$ (falsehood) and the binary
connectives $\Lolly$ (implication) and $\iAnd$ (conjunction).
We write $A\Lnot$ for $A \Lolly 1$ and $0$ for $1 \Lolly 1$.
Our choice of notation for connectives is that commonly used for
affine logic, since all the systems
we consider will turn out to be extensions of (minimal) affine logic.
Our use of $1$ rather than $0$ for falsehood is taken from continuous logic
\cite{ben-yaacov-pedersen09}, which motivated our work in this area.
In keeping
with this convention, we will order propositions by increasing logical strength,
so that $A \ge B$ means $A$ is at least as strong as $B$, i.e. that $A \Lolly B$ is provable.

As usual, we adopt the convention that $\Lolly$ associates to the right and has
lower precedence than $\iAnd$, which in turn has lower precedence than $(\cdot)\Lnot$.
So, for example, the brackets in $(A \iAnd  (B\Lnot)) \Lolly (C \Lolly (D \iAnd
F))$ are all redundant, while those in $(((A \Lolly B) \Lolly C) \iAnd D)\Lnot$
are all required.

\begin{figure}[h]
\begin{center}
\begin{tabular}{|cc|}
\hline
 & \\
\begin{prooftree}
\justifies
\Gamma, A \vdash A
\using{\ASM}
\end{prooftree}
&
\begin{prooftree}
\justifies
\Gamma, A \vdash A \iAnd A
\using{\CON}
\end{prooftree} \\
& \\
\begin{prooftree}
\justifies
\Gamma, 1 \vdash A
\using{\EFQ}
\end{prooftree} &
\begin{prooftree}
\justifies
\Gamma, \neg \neg A \vdash A
\using{\DNE}
\end{prooftree} \\ 
& \\
\multicolumn{2}{|c|}{%
\begin{prooftree}
\justifies
\Gamma, A, A \Lolly B \vdash B \iAnd (B \Lolly A)
\using{\CWC}
\end{prooftree} } \\ 
& \\
\hline
\end{tabular}
\caption{Sequent Calculus Axioms}
\label{fig:sequent-axioms}
\end{center}
\end{figure}
%

\subsection{The sequent calculi}

In this section we define the nine sequent calculi mentioned in the paper, though we will be mostly working
with the two fragments $\LLm$ and $\LLi$ of {\L}ukasiewicz logic. 
The judgments of the calculi are sequents $\Gamma \vdash A$ where the {\em
context} $\Gamma$ is a multiset of formulas and $A$ is a formula.
The rules of inference for \emph{all} the calculi
comprise the sequent formulation of a natural
deduction system shown in Figure~\ref{fig:sequent-rules}.

We will define the nine calculi by adding to the rules above some or all of the following axiom schemata:
\emph{assumption $\ASM$, contraction $\CON$}, {\it ex falso quodlibet} $\EFQ$,
\emph{double negation elimination} $\DNE$, and \emph{commutativity of weak conjunction} $\CWC$,
defined in Figure \ref{fig:sequent-axioms}. 
%
The nine calculi and their axiom schemata are as defined in the following table:
\begin{center}
\begin{tabular}{|c|l|} \hline
Calculus & Axiom Schemata \\ \hline
$\ALm$ & $\ASM$ \\ \hline
$\ALi$ & $\ASM$, $\EFQ$\\ \hline
$\ALc$ & $\ASM$, $\EFQ$, $\DNE$ \\ \hline
$\LLm$ & $\ASM$, $\CWC$ \\ \hline
$\LLi$ & $\ASM$, $\CWC$, $\EFQ$ \\ \hline
$\LLc$ & $\ASM$, $\CWC$, $\EFQ$, $\DNE$ \\ \hline
$\ML$ & $\ASM$, $\CON$ \\ \hline
$\IL$ & $\ASM$, $\CON$, $\EFQ$ \\ \hline
$\BL$ & $\ASM$, $\CON$, $\EFQ$, $\DNE$ \\ \hline
\end{tabular}
\end{center}
The systems $\ALm$, $\ALi$, $\ALc$, $\LLm$, $\LLi$ and $\LLc$ are
minimal, intuitionistic and classical variants
of affine logic and {\L}ukasiewicz logic.
$\ML$, $\IL$ and $\BL$ stand for minimal, intuitionistic and boolean logic for reasons to be made clear in Theorem \ref{thm-ml-il}.
The relationship between the nine logics is depicted in Figure~\ref{fig:logics}.


As our axiom schemata all allow additional premisses $\Gamma$ in the context, the following rule of weakening 
\[
\begin{prooftree}
\Gamma \vdash B
\justifies
\Gamma, A \vdash B
\using {\WK}
\end{prooftree}
\]
is admissible in all our logics, since given a proof tree with
$\Gamma \vdash B$ at the root, we may obtain a proof of $\Gamma, A \vdash B$
by adding $A$ to the context of every sequent on \emph{some} path from the root to a leaf (axiom). Also, note that in minimal affine logic $\ALm$, and hence
in all the logics,  the contraction axiom $\CON$ is inter-derivable with the contraction rule
\[
\begin{prooftree}
\Gamma, A, A \vdash B
\justifies
\Gamma, A \vdash B
\using {\CONR}
\end{prooftree}
\]
This proves the following theorem.

\begin{Theorem} \label{thm-ml-il} $\ML$, $\IL$ and $\BL$ are  the implication-conjunction fragments of the usual minimal, intuitionistic and boolean logics.
\Done
\end{Theorem}

Most of the results in this paper involve the derivability of a sequent in one of our calculi above. When deriving these, instead of writing proof trees we will typically adopt a form of equational reasoning, using the notations of the following definition.

\begin{Definition}
Let $T$ be an extension of $\ALm$. We write $A \ge_T B$ if $A \vdash B$ can be derived in $T$.
We write $A \simeq_T B$ if $A \ge_T B$ and $B \ge_T A$. When the $T$ in question is clear from the context we
just write $\ge$ and $\simeq$.
\end{Definition}

\begin{Lemma} \label{lma:ge-congruence} Let $T$ be an extension of $\ALm$.
Then $\le_T$ is symmetric and transitive, $\simeq_T$ is an equivalence
relation, and, for any formulas $A$, $B$ and $C$, such that $A \ge_T B$, the
following hold:
\begin{align*}
B \Lolly C &\ge_T A \Lolly C \\
C \Lolly A &\ge_T C \Lolly B  \\
A \iAnd C &\ge_T B \iAnd C \\
C \iAnd A &\ge_T C \iAnd B. 
\end{align*}
Hence, $\simeq$ is a congruence with respect to both $\Lolly$ and $\iAnd$,
i.e., if $A \simeq B$ then we have:
\begin{align*}
B \Lolly C &\simeq_T A \Lolly C \\
C \Lolly A &\simeq_T C \Lolly B  \\
A \iAnd C &\simeq_T B \iAnd C \\
C \iAnd A &\simeq_T C \iAnd B. 
\end{align*}
Moreover, we have:
\begin{align*}
A \iAnd (B \iAnd C) &\simeq_T (A \iAnd B) \iAnd C \\
A \iAnd B &\simeq_T B \iAnd A \\
A \iAnd 0 &\simeq_T A.
\end{align*}
and $A$ is provable iff $0 \ge_T A$ iff $A \simeq_T 0$.
\end{Lemma}
\Proof Straightforward, recalling that $0$ abbreviates $1 \Lolly 1$. \Done \\

When deriving an ``equation" such as $A \simeq_T B$ or an ``inequality" such as
$A \geq_T B$ we will often use the fact that $A \simeq A \iAnd 0$. For instance, assume we are given
some formula $B$ which is provable in $T$, so that $B \simeq_T 0$. We can then conclude $A \iAnd B \simeq_T A$. \\[2mm]
{\bf Notation}. Whenever, in a chain of equalities or inequalities, we make use of $B \simeq_T 0$ in the manner above, we will highlight the formula $B$ which (dis)appears 
by underlining it. Moreover, so that we can omit the subscript $T$ from $\leq_T$ and $\simeq_T$,
we will generally specify $T$ in brackets at the beginning of the statement of each lemma or theorem. The following lemma, which will be used later, illustrates both of these conventions.

\begin{Lemma}[$\LLm$] \label{lemma-cwc-lub} If $C \geq A$ and $C \geq B$ then $C \geq A \iAnd (A \Lolly B)$.
\end{Lemma}
\Proof Assume (a) $C \geq A$ (and hence $C \Lolly A \simeq 0$) and (b) $C \geq B$. We derive the conclusion as follows:
\begin{align*}
C
	& \simeq C \iAnd (\underline{C \Lolly A}) \by{a} \\
	& \simeq A \iAnd (A \Lolly C) \by{\CWC} \\
	& \geq A \iAnd (A \Lolly B) \by{b}
\end{align*}
invoking Lemma \ref{lma:ge-congruence} to use (b) in the last step.
\Done

\begin{Remark}
The rules of Figure~\ref{fig:sequent-rules}
and the axioms of Figure~\ref{fig:sequent-axioms}
are closed under substitution of formulas for variables.
Hence a substitution instance of a theorem in any of our logics is again
a theorem of that logic. When reading a result such as
Lemma~\ref{lemma-cwc-lub}, it is immaterial whether one views the letters $A$,
$B$ and $C$ as metavariables ranging over $\lL$ or as specific variables in
$\Var \subset \lL$.
\end{Remark}

\subsection{Hilbert-style systems}

Particularly in the literature on {\L}ukasiewicz logic, the
systems that we have presented as sequent calculi are traditionally
presented as Hilbert-style systems with {\it modus ponens} as the
only rule of inference. The following table defines the axiom
schemata {\em composition, commutativity of conjunction,
currying, uncurrying, weakening, {\it ex falso quodlibet},
double negation elimination, commutativity of weak
conjunction and contraction} that feature in these systems:
\[
\begin{array}{|c|l|} \hline
\axComp& (A \Lolly B) \Lolly (B \Lolly C) \Lolly (A \Lolly C) \\[0.5mm] \hline
\axComm& A \iAnd B \Lolly B \iAnd A \\[0.5mm] \hline
\axCurry& (A \iAnd  B \Lolly C) \Lolly (A \Lolly B \Lolly C) \\[0.5mm] \hline
\axUncurry& (A \Lolly B \Lolly C) \Lolly  (A \iAnd B \Lolly C) \\[0.5mm] \hline
\axWk& A \iAnd B \Lolly A \\[0.5mm] \hline
\axEFQ& 1 \Lolly A \\[0.5mm] \hline
\axDNE& A \Lnot\Lnot \Lolly A \\[0.5mm] \hline
\axCWC& A \iAnd (A \Lolly B) \Lolly B \iAnd (B \Lolly A) \\[0.5mm] \hline
\axCon& A \Lolly A \iAnd A \\[0.5mm] \hline
\end{array}
\]
We can then define nine Hilbert-style axiom systems as follows:
\[
\begin{array}{|c|l|} \hline
\Hilb{\ALm} & \axComp + \axComm + \axCurry + \axUncurry + \axWk \\[0.5mm] \hline
\Hilb{\ALi} & \Hilb{\ALm} + \axEFQ \\[0.5mm] \hline
\Hilb{\ALc} & \Hilb{\ALi} + \axDNE \\[0.5mm] \hline
\Hilb{\LLm} & \Hilb{\ALm} + \axCWC \\[0.5mm] \hline
\Hilb{\LLi} & \Hilb{\LLm} + \axEFQ \\[0.5mm] \hline
\Hilb{\LLc} & \Hilb{\LLi} + \axDNE \\[0.5mm] \hline
\Hilb{\ML}  & \Hilb{\ALm} + \axCon \\[0.5mm] \hline
\Hilb{\IL}  & \Hilb{\ML}  + \axEFQ \\[0.5mm] \hline
\Hilb{\BL}  & \Hilb{\IL}  + \axDNE \\[0.5mm] \hline
\end{array}
\]
The following theorem makes it precise the sense in which these are Hilbert-style versions of the nine sequent-style calculi presented in the previous section.

\begin{Theorem} For any formula $A \in \lL$, the sequent
\[ C_1, \ldots, C_k \vdash A \]
is provable in the sequent calculus
$\ALm$ (resp. 
$\ALi$,
$\ALc$,
$\LLm$,
$\LLi$,
$\LLc$,
$\ML$,
$\IL$,
$\BL$)
 iff the formula
\[  C_1 \Lolly C_2 \Lolly \ldots \Lolly C_k \Lolly A \] is derivable using {\it modus ponens} from the axiom systems $\Hilb{\ALm}$ (resp.
$\Hilb{\ALi}$,
$\Hilb{\ALc}$,
$\Hilb{\LLm}$,
$\Hilb{\LLi}$,
$\Hilb{\LLc}$,
$\Hilb{\ML}$,
$\Hilb{\IL}$ and
$\Hilb{\BL}$).
\end{Theorem}
\Proof
We sketch the proof for $\ALm$, the proofs for the other calculi
being straightforward extensions of this.
Let us write $A \Hilb{\ge} B$
if $A \Lolly B$ is derivable from  $\Hilb{\ALm}$ using {\it modus ponens}
and let us write $A \Hilb{\simeq} B$ if $A \Hilb{\ge} B$ and $B \Hilb{\ge} A$.
It is shown in~\cite{arthan-oliva14b} that $\Hilb{\ge}$ is a pre-order
and hence that $\Hilb{\simeq}$ is an equivalence relation and that the
equivalence classes of $\lL$ under $\Hilb{\simeq}$ form what
is known as a {\em pocrim}: an algebra $\VT$ comprising
{\em(i)} an ordered commutative monoid under
an operation $+$ induced by $\iAnd$ ($[A] + [B] = [A \iAnd B]$), with the equivalence class $[0]$
as the identity element ($[0] \iAnd [A] = [A]$) and least element ($[A] \ge [0]$) and
{\em(ii)} an operation $\rImp$ induced by $\Lolly$ ($[A] \rImp [B] = [A \Lolly B]$)
such that $A \Hilb{\ge} B$ iff $[A] \ge [B]$ iff
$[A] \Lolly [B] = [0]$ and such that the {\em residuation property} holds:
$[A] \iAnd [B] \ge [C]$ iff $[A] \ge [B] \Lolly [C]$.
It is then easy to see that $C_1, \ldots C_k \vdash A$ is provable in $\ALm$ iff
$[C_1] + \ldots + [C_k] \ge [A]$ in $\VT$ iff $C_1 \iAnd \ldots \iAnd C_k \Lolly A$ is derivable
from $\Hilb{\ALm}$ using {\em modus ponens}.
For example, the soundness of $\CE$ is equivalent to the property that $\gamma \ge a + b$
and $\delta + a + b \ge c$ implies $\gamma + \delta \ge c$ which holds in any ordered commutative monoid,
while the soundness of $\LI$ is equivalent to the residuation property.
\Done

\begin{Remark} \label{remark-rose}
{\L}ukasiewicz specified the logic that bears his name using
a standard semantics in which the truth values range over real numbers
in the interval $[0, 1]$ and conjectured that a certain Hilbert-style
system was complete for this semantics. A proof of this conjecture was apparently given in the 1930s by
Wajsberg but has been lost.
In the 1950s Rose and Rosser \cite{rose-rosser58}
(and also, by a rather more insightful method, Chang \cite{chang59})
proved an improved version of the conjecture, namely that the Hilbert-style
system $\Luk$ with {\it modus ponens} as the only inference rule and the
following axiom schemata is complete for the standard semantics.
\[
\begin{array}{rl}
\mbox{\textup{(A1)}} & A \Lolly (B \Lolly A) \\[1mm]
\mbox{\textup{(A2)}} & (A \Lolly B) \Lolly (B \Lolly C) \Lolly (A \Lolly C) \\[1mm]
\mbox{\textup{(A3)}} & ((A \Lolly B) \Lolly B) \Lolly ((B \Lolly A) \Lolly A) \\[1mm]
\mbox{\textup{(A4)}} & (A\Lnot \Lolly B\Lnot) \Lolly (B \Lolly A)
\end{array}
\]
In $\Luk$, implication and
negation are the primitive
connectives%
\footnote{Following {\L}ukasiewicz, Rose and Rosser used Polish notation,
writing $CAB$ for our $A \Lolly B$ and $NA$ for our $A\Lnot$,
as did Chang in the relatively few
fragments of syntax that appear in his treatment.}
Conjunction
$A \iAnd B$ can be defined in $\Luk$ as $(A \Lolly B\Lnot)\Lnot$. In \cite{arthan-oliva14b}, we show that the system $\Hilb{\LLc}$ is equivalent to $\Luk$,
justifying our identification of $\Hilb{\LLc}$ and hence of $\LLc$ with
(classical) {\L}ukasiewicz logic.
\end{Remark}

\Subsection{Derived connectives}
\label{sec-derived}

In additon to the primitive connectives $\iAnd$ and $\Lolly$, we will make extensive use of the following four \emph{derived} binary connectives
\[
\begin{array}{lcl}
	\CCb{A}{B} & \equiv & A \iAnd (A \Lolly B) \\[2mm]
	\CKb{A}{B} & \equiv & (B \Lolly A) \Lolly A \\[2mm]
	\CMb{A}{B} & \equiv & A \Lolly A \iAnd B \\[2mm]
	\oCKb{A}{B} & \equiv & A\Lnot \iAnd (B \Lolly A)
\end{array}
\]
Recall that we are assuming conjunction $\iAnd$ binds more strongly than the implication $\Lolly$, so that $\CMb{A}{B}$ is $A \Lolly (A \iAnd B)$. For the new connectives we will also use the convention that $\CC, \CK$ and $\oCK$ all bind more strongly than $\CM$. So $\CMb{(\CKb{A}{B})}{(\CCb{C}{D})}$, for instance, may be written as $\CMb{\CKb{A}{B}}{\CCb{C}{D}}$.

We justify our notation by observing that when $\iAnd$ and $\Lolly$ are replaced by the standard conjunction and implication of \emph{classical logic} then $\CCb{A}{B}, \CKb{A}{B}, \CMb{A}{B}$ are indeed equivalent to the standard conjunction, disjunction and implication. Moreover, $\oCKb{A}{B}$ is equivalent to the NOR binary connective. In affine linear logic, however, even the \emph{commutativity} of $\CCb{A}{B}$ and $\CKb{A}{B}$ are not derivable! The commutativity of $\CCb{A}{B}$ is the axiom schema $\CWC$ that defines $\LLi$, while the commutativity of $\CKb{A}{B}$ is axiom (A3) of Remark \ref{remark-rose}.

We have chosen our notation so that in each of the derived connectives the left operand appears both positively and negatively while the right operand appears only positively in $\CC, \CK$ and $\CM$ and only negatively in $\oCK$.
Monotonicity properties for the left operand of these connectives
have interesting connections with algebraic properties such
as commutativity and associativity. For example, for $\CCb{\relax}{\relax}$,
we have the following lemma:
\begin{Lemma}[$\ALm$] \label{monotone-second-op} Consider the following two axiom schema: the \emph{monotonicity of weak conjunction in its left operand}
\[
\begin{prooftree}
\justifies
\Gamma, A \Lolly B, \CCb{A}{C} \vdash \CCb{B}{C}
\using{\MWC}
\end{prooftree}
\]
and (one direction of) the \emph{associativity of weak conjunction}
\[
\begin{prooftree}
\justifies
\Gamma, \CCb{(\CCb{A}{B})}{C} \vdash \CCb{A}{(\CCb{B}{C})}
\using{\AWC}
\end{prooftree}
\]
Over $\ALm$ the axiom schema $\CWC$ is equivalent to either of these two.
\end{Lemma}
\Proof We have \\[1mm]
$\MWC$ implies $\CWC$: It is easy to check that if $X \geq Y$ then $X \simeq \CCb{X}{Y}$. 
Hence, taking $X$ to be $\CCb{A}{B}$ and $Y$ to be $A$, we have $\CCb{A}{B} \simeq \CCb{(\CCb{A}{B})}{A}$.
But using $\MWC$ then, as $\CCb{A}{B} \ge B$, we have $\CCb{(\CCb{A}{B})}{A} \ge \CCb{B}{A}$. \\[1mm]
$\CWC$ implies $\MWC$: This is clear since $\CCb{\relax}{\relax}$ is monotonic in its right operand. \\[1mm]
$\AWC$ implies $\CWC$: As above we have that $\CCb{A}{B} \simeq \CCb{(\CCb{A}{B})}{A}$. By $\AWC$ we have
$\CCb{(\CCb{A}{B})}{A} \geq \CCb{A}{(\CCb{B}{A})} \geq \CCb{B}{A}$. \\[1mm]
$\CWC$ implies $\AWC$: Easy using Lemma \ref{lemma-cwc-lub}, which says that, given $\CWC$, the conjunction $\CCb{A}{B}$ is a least upper bound of $A$ and $B$. \Done \\

Note that since, by Lemma~\ref{monotone-second-op},  $\AWC$ implies $\CWC$,
$\AWC$ also implies the other direction of the associativity of weak
conjunction: $\Gamma, \CCb{A}{(\CCb{B}{C})} \vdash  \CCb{(\CCb{A}{B})}{C}$.

\begin{Remark} A similar result to Lemma \ref{monotone-second-op} can be shown for $\CK$, i.e. over $\ALm$ the commutativity of $\CK$ is
equivalent to either its monotonicity of its left operand or its associativity. The operator $\oCK$ is not associative in any consistent extension of $\ALm$. 
However, over $\ALi$ it can be shown that $\oCK$ is anti-monotonic in its left operand iff it is commutative, and over $\ALc$, its commutativity is equivalent to $\CWC$.
\end{Remark}

We conclude this section with a short list of theorems of minimal affine logic $\ALm$ which will be used again and again in the proofs in Sections \ref{sec:weak-contraction} and \ref{sec-double-negation}.

\begin{Lemma} \label{ali-basic-lemma} The following are provable in $\ALm$
\begin{description}
	\item[$(i)$] $A \geq \CKb{B}{A}$
	\item[$(ii)$] $A \geq \CMb{B}{A}$
	\item[$(iii)$] $A \iAnd B \simeq A \iAnd (\CMb{A}{B})$
	\item[$(iv)$] $A \iAnd (B \Lolly C) \geq (A \Lolly B) \Lolly C$
	\item[$(v)$] $\CKb{C}{(A \Lolly B)} \geq (\CKb{C}{A}) \Lolly (\CKb{C}{B})$
	\item[$(vi)$] $\CKb{A}{B} \Lolly A \simeq B \Lolly A$.
\end{description}
\end{Lemma}
\Proof Easy. \Done

\Section{Minimal {\L}ukasiewicz Logic}
\label{sec:weak-contraction}

In this section we investigate minimal {\L}ukasiewicz logic $\LLm$, i.e.  we
investigate what is provable over affine logic using commutativity of weak
conjunction alone, with neither {\em ex falso quodlibet} nor double negation
elimination.  In $\LLm$, the constant $1$ has no special properties and could
be omitted.

We start the section by presenting an alternative axiomitisation of $\LLm$ over
$\ALm$, whereby instead of $\CWC$ we may take a restricted form of the contraction
axiom, which we term \emph{weak contraction}. We apply this to give a short
proof of a known result, Blok and Ferreirim's axiom L (Section
\ref{sec-axiom-l}). We then prove three important results about the derived
connectives $\CK, \CC$ and $\CM$ (Section \ref{sec-basic-k-c-m}). We conclude
with another application to a known result, a theorem of Ferreirim, Veroff and
Spinks (Section \ref{sec:f-v-s-theorem}), and a new result about idempotent formulas, i.e. formulas $A$ such that $A \simeq A \iAnd A$ (Section \ref{sec:idem}). 

\subsection{The weak contraction axiom $\WC$}
\label{weak-con-section}

Notice that even over $\ALm$ we have both
\[ A \vdash \CKb{B}{A} \quad \quad \mbox{and} \quad \quad A \vdash \CKb{B}{A} \Lolly A. \]
The first sequent is Lemma \ref{ali-basic-lemma} $(i)$, whereas the second is a straightforward consequence of weakening $\WK$. Hence, $\CKb{B}{A}$ and $\CKb{B}{A} \Lolly A$ are two immediate consequences of $A$. In this section we show that in order to capture {\L}ukasiewicz logic over affine logic we could have alternatively added to affine logic the following \emph{weak form of contraction}
\[ 
\begin{prooftree}
\justifies
\Gamma, A \vdash (\CKb{B}{A}) \iAnd (\CKb{B}{A} \Lolly A)
\using {\WC}
\end{prooftree} \]
Compare this with the standard contraction axiom $\Gamma, A \vdash A \iAnd A$. Rather than duplicating $A$, the weak contraction axiom says that we can replace $A$ by two weakenings of $A$. The converse of the implication is obvious, which means that $\WC$ implies (over $\ALm$)
\[ A \simeq (\CKb{B}{A}) \iAnd (\CKb{B}{A} \Lolly A). \] 

\begin{Remark} We note that the weak contraction axiom can also be nicely described via what we propose as the \emph{weak contraction rule}
\[
\begin{prooftree}
\Gamma, \CKb{B}{A} \Lolly A \vdash \CKb{B}{A} \Lolly C
\justifies
\Gamma, A \vdash C
\end{prooftree}
\]
It says that in order to prove that $A$ implies $C$, it is sufficient to prove that the weakening of $C$ by $\CKb{B}{A}$ follows from the weakening of $A$ by the same ``amount", and this holds for any choice of $B$.
\end{Remark}

\begin{Theorem} \label{luk-theorem}
$\LLm$ (resp. $\LLi$ and $\LLc$) is equivalent to $\ALm + \WC$ (resp. $\ALi + \WC$ and $\ALc + \WC$).
\end{Theorem}
\Proof We show that the axioms $\CWC$ and $\WC$ are interderivable over $\ALm$.
$\CWC$ may be derived from $\WC$ as follows:
\begin{align*}
A \iAnd (A \Lolly B)
	& \simeq (\CKb{B}{A}) \iAnd (\CKb{B}{A} \Lolly A) \iAnd (A \Lolly B) \by{\WC} \\
	& \geq (\CKb{B}{A}) \iAnd (B \Lolly A) \iAnd (A \Lolly B) \by{$B \geq \CKb{B}{A}$} \\
	& \simeq ((A \Lolly B) \Lolly B) \iAnd (B \Lolly A) \iAnd (A \Lolly B) \by{def $\CK$} \\
	& \geq B \iAnd (B \Lolly A). \by{easy}
\end{align*}
Conversely, in $\ALm + \CWC$ we have
\begin{align*}
A & \simeq A \iAnd (\underline{A \Lolly \CKb{B}{A}}) \by{Lemma \ref{ali-basic-lemma} $(i)$} \\
  & \simeq (\CKb{B}{A}) \iAnd (\CKb{B}{A} \Lolly A). \tag*{\CWC}
\end{align*}
As discussed above, it is easy to see that rule $\WC$ can be derived from $A \simeq (\CKb{B}{A}) \iAnd (\CKb{B}{A} \Lolly A)$.
\Done

\subsection{Application: Deriving axiom L}
\label{sec-axiom-l}


It is known that in {\L}ukasiewicz logic one can prove $\CKb{(A \Lolly B)}{(B
\Lolly A)}$. This is called axiom L by Blok and Ferreirim
\cite{blok-ferreirim00}, who give a proof due to Bosbach \cite{bosbach69a} of
an algebraic law equivalent to this result. Here we prove it using the weak
contraction axiom $\WC$.

\begin{Theorem}[$\LLm$] \label{axiom-l} $(B \Lolly A) \Lolly (A \Lolly B) \simeq A \Lolly B$
\end{Theorem}
\Proof $A \Lolly B \geq (B \Lolly A) \Lolly (A \Lolly B)$ follows directly by $\WK$. The other direction is derivable as
\begin{align*}
(B \Lolly A) \Lolly (A \Lolly B)
	& \simeq (B \Lolly A) \Lolly \CKb{B}{A} \Lolly B \by{Lemma \ref{ali-basic-lemma} ($vi$)} \\[1mm]
	& \simeq (\CKb{B}{A}) \iAnd (B \Lolly A) \Lolly B \by{easy} \\[1mm]
	& \geq (\CKb{B}{A}) \iAnd (\CKb{B}{A} \Lolly A) \Lolly B \by{$B \geq B \vee A$} \\[1mm]
	& \simeq A \Lolly B. \by{\WC}
\end{align*}
\Done \\

Hence it follows that \[ ((A \Lolly B) \Lolly (B \Lolly A)) \Lolly (B \Lolly A) = \CKb{(A \Lolly B)}{(B \Lolly A)} \simeq 0. \] 

\Subsection{Basic identities involving $\CK, \CC$ and $\CM$}
\label{sec-basic-k-c-m}

In this section we prove three basic identities that are valid in $\LLm$. The main identity is $A \iAnd B \simeq A \iAnd (\CKb{B}{(\CMb{A}{B})})$ which will be extensively used throughout the rest of the paper. The left-to-right implication is obvious, as $B \geq \CKb{B}{(\cdot)}$. The converse, however, says that in the context $A \iAnd (\cdot)$ we have $\CKb{B}{(\CMb{A}{B})} \geq B$, which is somewhat surprising.

\begin{Lemma}[$\LLm$] \label{k-cwc} $A \simeq (\CKb{A}{B}) \iAnd (B \Lolly A)$
\end{Lemma}
\Proof We have
\begin{align*}
A
	& \simeq A \iAnd \underline{(A \Lolly (B \Lolly A))} \by{\WK} \\[0mm]
	& \simeq ((B \Lolly A) \Lolly A) \iAnd (B \Lolly A) \by{\CWC} \\[1mm]
	& \simeq (\CKb{A}{B}) \iAnd (B \Lolly A). \by{\mbox{def $\CK$}}
\end{align*}
Recall that we underline easily proven conjuncts which are either inserted or deleted. \Done

\begin{Theorem}[$\LLm$] \label{lemma156} $A \iAnd B \simeq A \iAnd (\CKb{B}{(\CMb{A}{B})})$
\end{Theorem}
\Proof Let $X = ((\CMb{A}{B}) \Lolly B) \Lolly A$. We have
\begin{align*}
A \iAnd B
	& \simeq (\CMb{A}{B}) \iAnd A \by{Lemma \ref{ali-basic-lemma} $(iii)$} \\[1mm]
	& \simeq (\CMb{A}{B}) \iAnd A \iAnd \underline{(A \Lolly ((\CMb{A}{B}) \Lolly B))}  \by{easy} \\[0mm]
	& \simeq (\CMb{A}{B}) \iAnd ((\CMb{A}{B}) \Lolly B) \iAnd X \by{\CWC} \\[1mm]
	& \simeq B \iAnd \underline{(B \Lolly (\CMb{A}{B}))} \iAnd X \by{\CWC}  \\[0mm]
	& \simeq B \iAnd X \by{Lemma \ref{ali-basic-lemma} $(ii)$} \\[1mm]
	& \simeq (\CKb{B}{(\CMb{A}{B})}) \iAnd ((\CMb{A}{B}) \Lolly B) \iAnd X \by{Lemma \ref{k-cwc}} \\[1mm]
	& \simeq (\CKb{B}{(\CMb{A}{B})}) \iAnd A \iAnd \underline{(A \Lolly ((\CMb{A}{B}) \Lolly B))} \by{\CWC} \\[0mm]
	& \simeq A \iAnd (\CKb{B}{(\CMb{A}{B})}). \by{easy}
\end{align*}
\Done \\

Finally, we prove the following lemma which is used in Section \ref{sec:basic-logic}.

\begin{Lemma}[$\LLm$] \label{guess} $(A \Lolly C) \iAnd (C \Lolly B) \geq (A \Lolly B) \iAnd (\CCb{A}{B} \Lolly C)$
\end{Lemma}
\Proof We have
\begin{align*}
(A \Lolly C) \iAnd (C \Lolly B)
	& \geq (A \Lolly C) \iAnd ((A \Lolly C) \Lolly (A \Lolly B)) \by{easy} \\[1mm]
	& \simeq (A \Lolly B) \iAnd ((A \Lolly B) \Lolly (A \Lolly C)) \by{\CWC} \\[1mm]
	& \simeq (A \Lolly B) \iAnd (A \iAnd (A \Lolly B) \Lolly C) \by{easy} \\[1mm]
	& \simeq (A \Lolly B) \iAnd (\CCb{A}{B} \Lolly C). \by{def}
\end{align*}
\Done

\subsection{Application: The Ferreirim-Veroff-Spinks theorem}
\label{sec:f-v-s-theorem}

Ferreirim \cite{Ferreirim92} proved an algebraic formulation of the following
theorem, under extra assumptions, using model-theoretic methods.  With the
assistance of the Otter system \cite{McCune03} and Veroff's method of proof
sketches \cite{Veroff01}, Veroff and Spinks \cite{Veroff-Spinks04} found a
syntactic proof of the theorem in full generality.  An indirect proof
of the general result using algebraic methods is given in our companion
paper~\cite{arthan-oliva14b}.  Here we give a simplified and more abstract
version of the syntactic proof found by Veroff and Spinks, obtaining the result
as a straightforward consequence of the identity of Theorem~\ref{lemma156}.
(Our work in this area was in part inspired by a desire to understand the
Veroff-Spinks proof.)

\begin{Theorem}[$\LLm$] \label{veroff-proof} $$(A \Lolly A \otimes A) \iAnd (A \Lolly B \iAnd C) \geq (A \Lolly B) \iAnd (A \Lolly C)$$
\end{Theorem}
\Proof Let $X \equiv A \Lolly B \iAnd C$ and $Y = \CMb{(A \Lolly B)}{(A \Lolly C)}$. Note that
\begin{itemize}
	\item[] $(*) \; (A \Lolly A \iAnd A) \iAnd (A \Lolly (\CKb{(A \Lolly Z)}{W})) \geq \CKb{(A \Lolly Z)}{W}$
\end{itemize}
for any $Z$ and $W$. Hence
\begin{align*}
X	& \simeq X \iAnd (\underline{X \Lolly (A \Lolly B)}) \by{easy} \\[1mm]
	& \simeq (A \Lolly B) \iAnd ((A \Lolly B) \Lolly X) \by{$\CWC$} \\[1mm]
	& \geq (A \Lolly B) \iAnd ((A \Lolly B) \Lolly A \Lolly ((A \Lolly B) \iAnd (A \Lolly C))) \by{\WK} \\[1mm]
	& \simeq (A \Lolly B) \iAnd (A \Lolly Y) \by{def $\CM$} \\[1mm]
	& \geq (A \Lolly B) \iAnd (A \Lolly (\CKb{(A \Lolly C)}{Y})) \by{Lemma \ref{ali-basic-lemma} $(i)$} \\[1mm]
	& \geq (A \Lolly B) \iAnd (\CKb{(A \Lolly C)}{Y}) \by{$*$} \\[1mm]
	& \simeq (A \Lolly B) \iAnd (A \Lolly C). \by{Theorem \ref{lemma156}}
\end{align*}
\Done

\subsection{Application: Idempotent Sub-Hoops}
\label{sec:idem}

Let us call a formula $A$ \emph{idempotent} if $A \Lolly A \iAnd A$ is provable in $\LLm$. Recalling our abbreviation $\CMb{A}{B} = A \Lolly A \iAnd B$
the property of being idempotent can be written as $\CMb{A}{A}$. It is clear that the set of idempotent
formulas is closed under conjunction ($\iAnd$). In this section we show that it is also closed under implication ($\Lolly$).
This in particular implies that the set of idempotent elements of a hoop form a sub-hoop -- a result similar to that of Jipsen and Montagna \cite{jipsen-montagna06}
for the idempotent elements of a GBL algebra. We remark that although hoops are retracts of commutative GBL algebras, our proof does not seem
to be related to theirs since they make essential use the existence of joins in GBL algebras (while joins not necessarily exist in hoops).

\begin{Lemma}[$\LLm$] \label{idem-lemma} $\CKb{A}{(\CMb{A}{A})} \geq \CMb{A}{A}$.
\end{Lemma}
\Proof Immediate from Lemma \ref{lemma156}. \Done

\begin{Theorem}[$\LLm$] \label{idem-theorem} $(\CMb{A}{A}) \iAnd (\CMb{B}{B}) \Lolly (\CMb{(A \Lolly B)}{(A \Lolly B)})$.
\end{Theorem}
\Proof Asumme $(i) \; A \simeq A \iAnd A$ and $(ii) \; B \simeq B \iAnd B$. Abbreviating $B^A \equiv A \Lolly B$ we have
\begin{align*}
0
	& \simeq A \iAnd (A \Lolly B) \Lolly B \by{easy} \\[1mm]
	& \simeq A \iAnd (A \Lolly (A \Lolly B)) \Lolly B \by{$i$} \\[1mm]
	& \simeq A \iAnd (A \Lolly (B^A \Lolly B) \Lolly B)) \Lolly B \by{easy} \\[1mm]
	& \simeq A \iAnd (A \Lolly (B^A \Lolly B \iAnd B) \Lolly B)) \Lolly B \by{$ii$} \\[1mm]
	& \geq A \iAnd (A \Lolly (B^A \Lolly B^A \iAnd B^A) \Lolly B)) \Lolly B \by{easy} \\[1mm]
	& \simeq ((\CMb{B^A}{B^A}) \Lolly (A \Lolly B)) \Lolly (A \Lolly B) \by{easy} \\[1mm]
	& \geq \CMb{B^A}{B^A}. \by{Lemma \ref{idem-lemma}}
\end{align*}
\Done

\Section{Intuitionistic {\L}ukasiewicz Logic}
\label{sec-double-negation}

In this section, the constant $1$ starts to play a part:
we add {\em ex falso quodlibet}  to $\LLm$, giving rise to what we call
intuitionistic {\L}ukasiewicz logic $\LLi$. Recall that we
define negation by $A\Lnot = A \Lolly 1$.
Many of the results of this section attempt to unveil interesting
properties which are provable in $\LLi$ ``under" a negation. These include: some
basic properties of $\CK, \CM$ and $\CC$ (Section \ref{sec:basic-ili-k-m-c});
the duality between $\CK$ and $\oCK$ (Section \ref{sec:sym-k-ok}); 
homomorphism properties of double negation with respect to both implication (Section
\ref{homo-dn-lolly}) and conjunction (Section \ref{homo-dn-tensor}); and,
finally, a collection of ``De Morgan" properties (Section \ref{sec-de-morgan}).  

\Subsection{Basic identities on $\CK, \CM, \CC$ and negation}
\label{sec:basic-ili-k-m-c}

We start by establishing some basic identities about $\CK, \CM$ and $\CC$ in $\LLi$. The main result in this section is that the strong implication $\CM$ is a dual of a weak conjunction $\CC$ in the sense that $(\CCb{B}{A})\Lnot \simeq \CMb{A}{B\Lnot}$. This is akin to the relation between conjunction and implication $(A \iAnd B)\Lnot \simeq A \Lolly B\Lnot$ which one obtains in $\ALi$ simply by currying and uncurrying. 

\begin{Lemma}[$\LLi$] \label{lemma165} $A\Lnot \iAnd (\CKb{B}{A}) \simeq A\Lnot \iAnd B$
\end{Lemma}
\Proof The right-to-left direction follows directly from $B \geq \CKb{B}{A}$. For the other direction, note that by $\EFQ$ we have $A \geq \CMb{A\Lnot}{B}$. Hence, $\CKb{B}{A} \geq \CKb{B}{(\CMb{A\Lnot}{B})}$. Therefore, the result follows directly from Theorem \ref{lemma156}. \Done

\begin{Theorem}[$\LLi$] \label{thm-c-demorgan} $(\CCb{A}{B})\Lnot \simeq \CMb{A}{B\Lnot}$.
\end{Theorem}
\Proof Observe that by Lemma \ref{lemma165} (with $B$ and $A$ interchanged) it follows that $(*)$ $B\Lnot \geq \CKb{A}{B} \Lolly A \iAnd B\Lnot$. Hence
\begin{align*}
(\CCb{A}{B})\Lnot
	& \simeq (A \iAnd (A \Lolly B))\Lnot \by{\mbox{def $\CC$}} \\[1mm]
	& \simeq (B \iAnd (B \Lolly A))\Lnot \by{\mbox{$\CWC$}} \\[1mm]
	& \simeq (B \Lolly A) \Lolly B\Lnot \by{\mbox{easy}} \\[1mm]
	& \geq (B \Lolly A) \Lolly (\CKb{A}{B}) \Lolly A \iAnd B\Lnot \by{$*$} \\[1mm]
	& \simeq A \Lolly (\underline{A \Lolly (B \Lolly A)}) \Lolly A \iAnd B\Lnot \by{\CWC} \\[1mm]
	& \simeq A \Lolly A \iAnd B\Lnot. \by{\WK} \\[1mm]
	& \simeq \CMb{A}{B\Lnot}. \by{def $\CM$}
\end{align*}
The converse implication is straightforward. \Done

\begin{Corollary}[$\LLi$] \label{lemma48459} $\CMb{A}{B\Lnot} \simeq \CMb{B}{A\Lnot}$
\end{Corollary}
\Proof Direct from Theorem \ref{thm-c-demorgan}, since $\CC$ is commutative (i.e. $\CWC$). \Done

\Subsection{Symmetries of $\CK$ and $\oCK$ and $\DNE$}
\label{sec:sym-k-ok}

Although the commutativity of $\CK$ is clearly a classical principle, it is perhaps surprising that commutativity of $\oCKb{B}{A} $ can be proven intuitionistically.

\begin{Theorem}[$\LLi$] \label{lemma-a-lolly-b-not-a} $\oCKb{B}{A} \simeq \oCKb{A}{B}$
\end{Theorem}
\Proof By symmetry it is enough to prove $\oCKb{B}{A} \geq \oCKb{A}{B}$. We have
\begin{align*}
(B \Lolly A) \iAnd A\Lnot & \simeq (B \Lolly A) \iAnd A\Lnot \iAnd \underline{(A\Lnot \Lolly (A \Lolly B))} \by{\EFQ} \\[0mm]
	& \simeq (B \Lolly A) \iAnd (A \Lolly B) \iAnd ((A \Lolly B) \Lolly A\Lnot) \by{\CWC} \\[1mm]
	& \simeq (B \Lolly A) \iAnd (A \Lolly B) \iAnd ((A \Lolly B) \iAnd A)\Lnot \by{easy} \\[1mm]
	& \simeq (B \Lolly A) \iAnd (A \Lolly B) \iAnd (B \iAnd (B \Lolly A))\Lnot \by{\CWC} \\[1mm]
	& \simeq (A \Lolly B) \iAnd (B \Lolly A) \iAnd ((B \Lolly A) \Lolly B\Lnot) \by{easy} \\[1mm]
	& \geq (A \Lolly B) \iAnd B\Lnot. \by{easy}
\end{align*}
Recall that we underline easily proven conjuncts which are either inserted or deleted. \Done \\

A surprising consequence of this is that the double negation of the classical axiom $\DNE$ is provable intuitionistically, i.e. in $\LLi$. It is well known that this is true for full intuitionistic logic\footnote{In $\IL$ the proof goes as follows: Assuming (1) $(A\Lnot\Lnot \Lolly A)\Lnot$ we must derive a contraction. First use (1) to derive $A\Lnot$, by $\WK$. Assume also (2) $A\Lnot\Lnot$. From (2) and $A\Lnot$ we obtain $1$, and hence $A$. Hence, discharging the assumption (2) we have $A\Lnot\Lnot \Lolly A$, which by (1) gives a contradiction. Note, however, that assumption (1) was used twice.} $\IL$, but that proof makes apparently essential use of the full contraction axiom. That the result can be proved using only the weak form of contraction permitted by $\CWC$ is rather amazing.

\begin{Corollary}[$\LLi$] \label{thm139} $(A\Lnot\Lnot \Lolly A)\Lnot\Lnot$
\end{Corollary}
\Proof Note that, since $1 \simeq A \iAnd A\Lnot$ we have $(*) \; A\Lnot\Lnot \simeq \CMb{A\Lnot}{A}$. Moreover, it is easy to check that $(**) \; \oCKb{X}{(Y \Lolly X)} \simeq X\Lnot \iAnd (\CKb{X}{Y})$, for all $X$ and $Y$. Hence
\begin{align*}
(A\Lnot\Lnot \Lolly A)\Lnot
	& \simeq ((\CMb{A\Lnot}{A}) \Lolly A)\Lnot \by{$*$} \\[1mm]
	& \simeq ((\CMb{A\Lnot}{A}) \Lolly A)\Lnot \iAnd \underline{(A \Lolly ((\CMb{A\Lnot}{A}) \Lolly A))} \by{\WK} \\[1mm]
	& \simeq \oCKb{((\CMb{A\Lnot}{A}) \Lolly A)}{A} \by{def $\oCK$} \\[1mm]
	& \simeq \oCKb{A}{((\CMb{A\Lnot}{A}) \Lolly A)} \by{Theorem \ref{lemma-a-lolly-b-not-a}} \\[1mm]
	& \simeq A\Lnot \iAnd (\CKb{A}{(\CMb{A\Lnot}{A})}) \by{$**$} \\[1mm]
	& \simeq A\Lnot \iAnd A \by{Lemma \ref{lemma156}} \\[1mm]
	& \simeq 1. \tag*{\Done} \\[-8mm]
\end{align*} \\[-2mm]

Finally, the following theorem shows that the NOR connective $\oCK$ is indeed the negation of the disjunction $\CK$, a fact which holds in full intuitionistic logic $\IL$, but again, the simplest proof seems to make essential use of the full contraction axiom. 

\begin{Theorem}[$\LLi$] \label{lemma-k-negated} $(\CKb{A}{B})\Lnot \simeq \oCKb{A}{B}$
\end{Theorem}
\Proof By Lemma \ref{ali-basic-lemma} $(i)$ we have $B \Lolly \CKb{A}{B}$; and by $\EFQ$ we have $1 \Lolly A$. Hence, $(*) \; (\CKb{A}{B})\Lnot \Lolly (B \Lolly A)$. Therefore
\begin{align*}
(\CKb{A}{B})\Lnot & \simeq (\CKb{A}{B})\Lnot \iAnd ((\CKb{A}{B})\Lnot \Lolly (B \Lolly A)) \by{$*$} \\[1mm]
	& \simeq (B \Lolly A) \iAnd ((B \Lolly A) \Lolly (\CKb{A}{B})\Lnot) \by{\CWC} \\[1mm]
	& \simeq (B \Lolly A) \iAnd ((B \Lolly A) \iAnd (\CKb{A}{B}))\Lnot \by{easy} \\[1mm]
	& \simeq (B \Lolly A) \iAnd (A \iAnd (\underline{A \Lolly (B \Lolly A)}))\Lnot \by{\CWC} \\[1mm]
	& \simeq (B \Lolly A) \iAnd A\Lnot. \by{\WK}
\end{align*}
\Done \\

Finally, the results above imply the commutativity of $\CK$ under a negation.

\begin{Theorem}[$\LLi$] $(\CKb{A}{B})\Lnot \simeq (\CKb{B}{A})\Lnot$
\end{Theorem}
\Proof Direct from Theorems \ref{lemma-a-lolly-b-not-a} and \ref{lemma-k-negated}. \Done

\subsection{Double negation homomorphism: Implication}
\label{homo-dn-lolly}

We now show (in $\LLi$) that the double negation operation $(\cdot)\Lnot\Lnot$ is a homomorphism for implication, i.e.
\[ (A \Lolly B)\Lnot\Lnot \simeq A\Lnot\Lnot \Lolly B\Lnot\Lnot. \]
We will show the same for conjunction in Section \ref{homo-dn-tensor}.

Note that $\CKb{1}{A} = A\Lnot\Lnot$. Hence, it follows from Lemma \ref{ali-basic-lemma} ($v$) that $(A \Lolly B)\Lnot\Lnot \geq A\Lnot\Lnot \Lolly B\Lnot\Lnot$ and hence $(A \Lolly B)\Lnot\Lnot \geq A \Lolly B\Lnot\Lnot$ is provable already in $\ALm$, i.e. without making use of $\EFQ$. In this section we show that the converse implications hold in $\LLi$. Again, the fact that this holds in full intuitionistic logic is well known. See \cite{Troelstra(73)}, page 9, for instance, for an $\IL$-derivation of Theorem \ref{thm-lolly-not-not}. That derivation, however, uses the assumption $(A \Lolly B)\Lnot$ twice, and hence cannot be formalised in $\LLi$. 

\begin{Theorem}[$\LLi$] \label{thm-lolly-not-not} $A\Lnot\Lnot \Lolly B\Lnot\Lnot \geq (A \Lolly B)\Lnot\Lnot$
\end{Theorem}
\Proof We have
\begin{align*}
A\Lnot\Lnot \Lolly B\Lnot\Lnot
	& \geq A \Lolly B\Lnot\Lnot \by{$A \geq A\Lnot\Lnot$} \\[1mm]
	& \geq (B\Lnot\Lnot \Lolly B) \Lolly (A \Lolly B) \by{easy} \\[1mm]
	& \geq (A \Lolly B)\Lnot \Lolly (B\Lnot\Lnot \Lolly B)\Lnot \by{easy} \\[1mm]
	& \simeq (A \Lolly B)\Lnot \Lolly 1 \by{Corollary \ref{thm139}} \\[1mm]
	& \simeq (A \Lolly B)\Lnot\Lnot. \by{def $(\cdot)\Lnot$}
\end{align*}
\Done


\subsection{Application: Basic Logic}
\label{sec:basic-logic}

We have seen in Section \ref{sec-axiom-l} that in minimal {\L}ukasiewicz logic $\LLm$ we can prove the disjunction $\CKb{(A \Lolly B)}{(B \Lolly A)}$. It is well known that in $\LLc$ one can also prove the following (stronger) result\footnote{This result does not hold intuitionistically. Consider the Kripke structure with three nodes $w_0, w_1, w_2$ with $w_0 < w1,w2$ and $w_0 = \{ \, \}, w_1 = \{P, R\}, w_2 = \{Q, R\}$. Although both $(P \Lolly Q) \Lolly R$ and $(Q \Lolly P) \Lolly R$ hold in $w_0$,  $R$ fails to hold at $w_0$.}
\[ (A \Lolly B) \Lolly C, (B \Lolly A) \Lolly C \vdash C \]
which is axiom (A6) of \emph{basic logic} \cite{Hajek98}. However, it is easy to construct a counter-example (bounded hoop) showing that (A6) is not provable in $\LLi$ in general. Nevertheless, we show that this can be derived in $\LLi$ plus one single application of $\DNE$, i.e. we show that the following sequent is provable in $\LLi$:
\[ (A \Lolly B) \Lolly C, (B \Lolly A) \Lolly C \vdash C\Lnot\Lnot. \]

\begin{Lemma}[$\LLi$] \label{guess-il} $A\Lnot \simeq (A \Lolly B) \iAnd (\CCb{A}{B})\Lnot$.
\end{Lemma}
\Proof Left-to-right follows directly from Lemma \ref{guess}, taking $C = 1$. For the converse observe that $(\CCb{A}{B})\Lnot \simeq A \Lolly (A \Lolly B)\Lnot$. \Done

\begin{Theorem}[$\LLi$] \label{cor-b6} $(A \Lolly B) \Lolly C, (B \Lolly A) \Lolly C \vdash C\Lnot\Lnot$
\end{Theorem}
\Proof We will prove equivalently that the following holds:
\[ ((A \Lolly B) \Lolly C) \iAnd ((B \Lolly A) \Lolly C) \iAnd C\Lnot \simeq 1.\]
Let $X = (A \Lolly B) \Lolly C$ and $Y = (B \Lolly A) \Lolly C$. We have
\begin{align*}
X \iAnd Y \iAnd C\Lnot
	& \simeq X \iAnd Y \iAnd (C \Lolly (B \Lolly A)) \iAnd (\CCb{C}{(B \Lolly A)})\Lnot \by{Lemma \ref{guess-il}} \\[1mm]
	& \simeq X \iAnd Y \iAnd (C \Lolly (B \Lolly A)) \iAnd (\CCb{(B \Lolly A)}{C})\Lnot \by{\CWC} \\[1mm]
	& \geq Y \iAnd ((A \Lolly B) \Lolly (B \Lolly A)) \iAnd (Y \Lolly (B \Lolly A)\Lnot) \by{easy} \\[1mm]
	& \simeq Y \iAnd (B \Lolly A) \iAnd (Y \Lolly (B \Lolly A)\Lnot) \by{Theorem \ref{axiom-l}} \\[1mm]
	& \geq (B \Lolly A) \iAnd (B \Lolly A)\Lnot \by{easy} \\[1mm]
	& \simeq 1. \by{easy}
\end{align*}
\Done


\subsection{Double negation homomorphism: Conjunction}
\label{homo-dn-tensor}


As done in Section \ref{homo-dn-lolly} for implication, we now show that (in $\LLi$) the double negation operation $(\cdot)\Lnot\Lnot$ is also a homomorphism for conjunction, i.e.
\[ (A \iAnd B)\Lnot\Lnot \simeq A\Lnot\Lnot \iAnd B\Lnot\Lnot. \]
This result will follow immediately from the following surprising duality between implication ($\Lolly$) and conjunction ($\iAnd$).

\begin{Theorem}[$\LLi$] \label{lolly-plus} $(A\Lnot \Lolly B)\Lnot \simeq A\Lnot \iAnd B\Lnot$
\end{Theorem}
\Proof The implication from right to left is easy. Since, by $\EFQ$ we have $A\Lnot \Lolly 1 \ge A\Lnot \Lolly B$, we obtain
\[ (*) \; A\Lnot\Lnot \Lolly (A\Lnot\Lnot \iAnd (A\Lnot \Lolly B)\Lnot) \ge A\Lnot\Lnot \Lolly (A\Lnot\Lnot \iAnd A\Lnot\Lnot\Lnot) \simeq A\Lnot\Lnot\Lnot. \]
Hence, taking $A' = A\Lnot \Lolly B$ and $B' = A\Lnot\Lnot$ in Lemma \ref{guess-il}, we have the first line of the following chain
\begin{align*}
(A\Lnot \Lolly B)\Lnot
	& \simeq ((A\Lnot \Lolly B) \Lolly A\Lnot\Lnot) \iAnd (\CCb{(A\Lnot \Lolly B)}{A\Lnot\Lnot})\Lnot \\[1mm]	
	& \simeq ((A\Lnot \Lolly B) \Lolly A\Lnot\Lnot) \iAnd (\CMb{A\Lnot\Lnot}{(A\Lnot \Lolly B)\Lnot}) \by{Theorem \ref{thm-c-demorgan}} \\[1mm]	
	& \geq ((A\Lnot \Lolly B) \Lolly A\Lnot\Lnot) \iAnd A\Lnot\Lnot\Lnot \by{$*$} \\[1mm]
	& \simeq (A\Lnot \iAnd (A\Lnot \Lolly B))\Lnot \iAnd A\Lnot \by{easy} \\[1mm]
	& \simeq (\CCb{A\Lnot}{B})\Lnot \iAnd A\Lnot \by{def $\CC$} \\[1mm]
	& \simeq (\CMb{B}{A\Lnot\Lnot}) \iAnd A\Lnot \by{Theorem \ref{thm-c-demorgan}} \\[1mm]
	& \simeq (\CMb{A\Lnot}{B\Lnot}) \iAnd A\Lnot \by{Corollary \ref{lemma48459} $(i)$} \\[1mm]
	& \geq A\Lnot \iAnd B\Lnot. \by{easy}
\end{align*}
\Done

\begin{Theorem}[$\LLi$] \label{main-thm-n-trans} $(A \iAnd B)\Lnot\Lnot \simeq A\Lnot\Lnot \iAnd B\Lnot\Lnot$
\end{Theorem}
\Proof By Theorem \ref{lolly-plus}, since $(A \iAnd B)\Lnot\Lnot \simeq (A\Lnot\Lnot \Lolly B\Lnot)\Lnot$. \Done

\Subsection{Some De Morgan properties}
\label{sec-de-morgan}

Let us conclude this section on $\LLi$ with a list of De Morgan laws for the all our connectives (primitive and derived). For conjunction ($\iAnd$) this is a trivial consequence of (un)currying, whereas for the weak conjunction ($\CC$) this is shown in Theorem \ref{thm-c-demorgan}. We prove here similar results for the other connectives.

\begin{Theorem} \label{thm-main-demorgan} The following ``De Morgan dualities" hold in intuitionistic {\L}ukasiewicz logic $\LLi$
\[
\begin{array}{rcl}
(A \iAnd B)\Lnot & \simeq & A \Lolly B\Lnot \\[2mm]
(A \Lolly B)\Lnot & \simeq & A\Lnot\Lnot \iAnd B\Lnot \\[2mm]
(\CCb{A}{B})\Lnot & \simeq & \CMb{A}{B\Lnot} \\[2mm]
(\CMb{A}{B})\Lnot & \simeq & \CCb{A\Lnot\Lnot}{B\Lnot} \\[2mm]
(\CCb{A}{B})\Lnot & \simeq & \CKb{A\Lnot}{B\Lnot} \\[2mm]
(\CKb{A}{B})\Lnot & \simeq & \CCb{A\Lnot}{B\Lnot} \\[2mm]
(\oCKb{A}{B})\Lnot & \simeq & \CMb{A\Lnot}{B\Lnot\Lnot}.
\end{array}
\]
\end{Theorem}
\Proof The first equation $(A \iAnd B)\Lnot \simeq A \Lolly B\Lnot$ follows directly from currying and uncurrying. For the second equation we calculate as follows
\begin{align*}
(A \Lolly B)\Lnot & \simeq (A \Lolly B)\Lnot\Lnot\Lnot \by{easy} \\[1mm]
	& \simeq (A\Lnot\Lnot \Lolly B\Lnot\Lnot)\Lnot \by{Theorem \ref{thm-lolly-not-not}} \\[1mm]
	& \simeq (B\Lnot \Lolly A\Lnot)\Lnot \by{easy} \\[1mm]
	& \simeq A\Lnot\Lnot \iAnd B\Lnot. \by{Theorem \ref{lolly-plus}}
\end{align*}
The third equation follows from Theorem \ref{thm-c-demorgan} and $\CWC$. The fourth equation can be derived as:
\begin{align*}
(\CMb{A}{B})\Lnot
	& \simeq (A \Lolly A \iAnd B)\Lnot \by{def $\CM$} \\[1mm]
	& \simeq A\Lnot\Lnot \iAnd (A \iAnd B)\Lnot \by{duality of $\Lolly$} \\[1mm]
	& \simeq A\Lnot\Lnot \iAnd (B \Lolly A\Lnot) \by{duality of $\iAnd$} \\[1mm]
	& \simeq A\Lnot\Lnot \iAnd (B \Lolly A\Lnot\Lnot\Lnot) \by{$A\Lnot \simeq A\Lnot\Lnot\Lnot$} \\[1mm]
	& \simeq A\Lnot\Lnot \iAnd (A\Lnot\Lnot \Lolly B\Lnot) \by{easy} \\[1mm]
	& \simeq \CCb{A\Lnot\Lnot}{B\Lnot}. \by{def $\CC$}
\end{align*}
The fifth equation follows by:
\begin{align*}
(\CCb{A}{B})\Lnot
	& \simeq (A \iAnd (A \Lolly B))\Lnot \by{easy} \\[1mm]
	& \simeq (A \Lolly B)\Lnot\Lnot \Lolly A\Lnot \by{easy} \\[1mm]
	& \simeq (A\Lnot\Lnot \Lolly B\Lnot\Lnot) \Lolly A\Lnot \by{Theorems \ref{thm-lolly-not-not}} \\[1mm]
	& \simeq (B\Lnot \Lolly A\Lnot) \Lolly A\Lnot \by{easy} \\[1mm]
	& \simeq \CKb{A\Lnot}{B\Lnot}.
\end{align*}
For the sixth equation we proceed as follows:
\begin{align*}
(\CKb{B}{A})\Lnot
	& \simeq ((A \Lolly B) \Lolly B)\Lnot \by{def $\CK$} \\[1mm]
	& \simeq (A \Lolly B)\Lnot\Lnot \iAnd B\Lnot \by{duality of $\Lolly$} \\[1mm]
	& \simeq (A\Lnot\Lnot \iAnd B\Lnot)\Lnot \iAnd B\Lnot \by{duality of $\Lolly$} \\[1mm]
	& \simeq (B\Lnot \Lolly A\Lnot\Lnot\Lnot) \iAnd B\Lnot \by{duality of $\iAnd$} \\[1mm]
	& \simeq (B\Lnot \Lolly A\Lnot) \iAnd B\Lnot \by{$A\Lnot \simeq A\Lnot\Lnot\Lnot$} \\[1mm]
	& \simeq \CCb{B\Lnot}{A\Lnot}. \by{def $\CC$}
\end{align*}
Finally, the last equation follows from Theorem \ref{lemma-k-negated} and the laws for $\CC$ and $\CK$. \Done

\begin{Corollary}[$\LLi$] $\oCKb{A}{B} \simeq \CCb{A\Lnot}{B\Lnot}$
\end{Corollary}
\Proof Immediate from Theorem \ref{lemma-k-negated} and the last identity of Theorem \ref{thm-main-demorgan} above. \Done \\

Unfolding the definitions of $\oCK$ and $\CC$ the corollary says 
\[ A\Lnot \iAnd (B \Lolly A) \simeq A\Lnot \iAnd (A\Lnot \Lolly B\Lnot) \]
which means that \emph{in the context} $A\Lnot \iAnd (\cdot)$ the two implications $B \Lolly A$ and $A\Lnot \Lolly B\Lnot$ are intuitionistically equivalent.

\subsection{Application: $k$-contradiction implies $\DNE$}

The sequent $A\Lnot, A\Lnot\Lnot \vdash A$ is obviously provable.
Model-theoretic considerations suggest that this sequent is the first in a
sequence of provable sequents in which the hypothesis $A\Lnot$ is weakened to
$(A \iAnd A)\Lnot$, $(A \iAnd A \iAnd A)\Lnot$, \ldots.
See~\cite{arthan-oliva14b} for further discussion and a model-theoretic proof.
Finding syntactic proofs of these sequents is an interesting application of
many of the results we have proven so far.

\begin{Theorem}[$\LLi$] \label{theorem-k-contradiction}
Whenever $k \geq 1$ copies of $A$ lead to a contradiction, then we must have $\DNE$ for $A$, i.e.
\[ (\underbrace{A \iAnd \ldots \iAnd A}_{\mbox{$k$ times}})\Lnot, A\Lnot\Lnot \vdash A. \]
\end{Theorem}
\Proof By induction on $k$. The base case $k = 1$ is trivial, i.e. $A\Lnot, A\Lnot\Lnot \vdash A$. 
For the induction hypothesis we will show that the following rule is derivable for $k \geq 1$.
\[ 
\begin{prooftree}
	(A^k)\Lnot, A\Lnot\Lnot \vdash A
	\Justifies
	(A \iAnd A^k)\Lnot, A\Lnot\Lnot \vdash A
\end{prooftree}
\]
In fact, we show the much stronger result that the following rule is derivable:
\[ 
\begin{prooftree}
	B\Lnot, A\Lnot\Lnot \vdash A
	\Justifies
	(A \iAnd B)\Lnot, A\Lnot\Lnot \vdash A
\end{prooftree}
\]
The derivation is immediate from Lemma \ref{lemma-k-contradiction} below. \Done

\begin{Lemma}[$\LLi$] \label{lemma-k-contradiction} $B\Lnot \Lolly A\Lnot\Lnot \Lolly A \geq (A \iAnd B)\Lnot \Lolly A\Lnot\Lnot \Lolly A$.
\end{Lemma}
\Proof Let $X \equiv B\Lnot \Lolly A\Lnot\Lnot \Lolly A$ and $Y \equiv (A \iAnd B)\Lnot \iAnd A\Lnot\Lnot$. We have 
\begin{align*}
X \iAnd Y & \simeq (A\Lnot\Lnot \Lolly B\Lnot \Lolly A) \iAnd A\Lnot\Lnot \iAnd (A \Lolly B\Lnot) \by{easy} \\[1mm]
	& \simeq (B\Lnot \Lolly A) \iAnd ((B\Lnot \Lolly A) \Lolly A\Lnot\Lnot) \iAnd (A \Lolly B\Lnot) \by{\CWC} \\[1mm]
	& \simeq (B\Lnot \Lolly A) \iAnd (A\Lnot \Lolly (B\Lnot \Lolly A)\Lnot) \iAnd (A \Lolly B\Lnot) \by{easy} \\[1mm]
	& \simeq (B\Lnot \Lolly A) \iAnd (A\Lnot \Lolly A\Lnot \iAnd B\Lnot) \iAnd (A \Lolly B\Lnot) \by{Theorem \ref{lolly-plus}} \\[1mm]
	& \simeq (B\Lnot \Lolly A) \iAnd (\CMb{A\Lnot}{B\Lnot}) \iAnd (A \Lolly B\Lnot) \by{def $\CM$} \\[1mm]
	& \simeq (B\Lnot \Lolly A) \iAnd (\CCb{B}{A\Lnot})\Lnot \iAnd (A \Lolly B\Lnot) \by{Theorem \ref{thm-main-demorgan}} \\[1mm]
	& \simeq (B\Lnot \Lolly A) \iAnd ((A \iAnd B)\Lnot \Lolly B\Lnot) \iAnd (A \iAnd B)\Lnot \by{easy} \\[1mm]
	& \simeq (B\Lnot \Lolly A) \iAnd B\Lnot \iAnd \underline{(B\Lnot \Lolly (A \iAnd B)\Lnot)} \by{\CWC} \\[0mm]
	& \simeq (B\Lnot \Lolly A) \iAnd B\Lnot. \by{easy} \\[1mm]
	& \geq A. \by{easy}
\end{align*}
\Done

%

\Section{Classical {\L}ukasiewicz Logic}
\label{sec:embedding}

In this section we investigate how the well-known double negation translations \cite{FO(2012B)} of classical (Boolean) logic $\BL$ into $\IL$ map over to the setting of {\L}ukasiewicz Logic. Our starting assumption was that due to the widespread use of contraction in the proof of soundness for these translations, we would find at least one double negation translation of $\BL$ into $\IL$ which would fail as a translation of $\LLc$ into $\LLi$. Therefore, we were amazed to discover that in fact all of the well-known double negation translations of classical logic also work in the presence of the weak contraction available in {\L}ukasiewicz Logic. This is a non-trivial result, since as we will also shown, some of these translation do fail when no contraction is available, as for instance in affine logic. The crucial property needed here is that the double negation mapping $A \mapsto A\Lnot\Lnot$ is a homomorphism, as shown in Sections \ref{homo-dn-lolly} and \ref{homo-dn-tensor}.

\Subsection{Double negation translations}

We adapt Troelstra's definition (cf. \cite{Troelstra(73)}, section 10) which imposes three requirements on a double negation translation.


\begin{Definition} Let ${\bf A}$ be an extension of $\ALi$. A formula translation $(\cdot)^\dagger \colon {\cal L} \to {\cal L}$ is a \emph{double negation translation for {\bf A}} if the following hold for every formula $A$ in the language of ${\bf A}$ 
\begin{description}
	\item[(DNS1)] ${\bf A} + \DNE$ proves $A^\dagger \vdash A$ and $A \vdash A^\dagger$.
	\item[(DNS2)] if ${\bf A} + \DNE$ proves $\vdash A$ then ${\bf A}$ proves $\vdash A^\dagger$.
	\item[(DNS3)] ${\bf A}$ proves\footnote{Troestra's third condition (DNS3) is slightly different in that it requires $A^\dagger$ to be built of double negated atoms from ``negative connectives" (so as to rule out existential quantifiers and disjunction). In our setting where affine linear logic is the starting point, this complication does not arise.} $(A^\dagger)\Lnot\Lnot \vdash A^\dagger$.
\end{description}
\end{Definition}

\newcommand{\kTrans}[1]{{#1}^{\sf K}}

First of all, we show that both the Kolmogorov and the G\"odel translations are in fact double negation translations for \emph{affine} logic, i.e. no contraction is necessary to prove Troelstra's three requirements. Let ${\cal L}$ be the language of the theories $\ALc$ and $\ALi$.

\begin{Definition}[Kolmogorov translation \cite{Kolmogorov(25)}] For each formula $A \in {\cal L}$ associate a formula $\kTrans{A} \in {\cal L}$ inductively as follows:
\[
\begin{array}{rcl}
	\kTrans{P} & \equiv & P\Lnot\Lnot \quad\quad (\mbox{$P$ atomic}) \\[2mm]
	\kTrans{1} & \equiv & 1 \\[2mm]
	\kTrans{(A \iAnd B)} & \equiv & (\kTrans{A} \iAnd \kTrans{B})\Lnot\Lnot \\[2mm]
	\kTrans{(A \Lolly B)} & \equiv & (\kTrans{A} \Lolly \kTrans{B})\Lnot\Lnot.
\end{array}
\]
\end{Definition}

\newcommand{\godTrans}[1]{{#1}^{\mbox{{\scriptsize \sf G\"{o}}}}}

\begin{Definition}[G\"odel translation \cite{goedel33a}] For each formula $A \in {\cal L}$ associate a formula $\godTrans{A} \in {\cal L}$ inductively as follows:
\[
\begin{array}{rcl}
	\godTrans{P} & \equiv & P \quad\quad (\mbox{$P$ atomic}) \\[2mm]
	\godTrans{1} & \equiv & 1 \\[2mm]
	\godTrans{(A \iAnd B)} & \equiv & \godTrans{A} \iAnd \godTrans{B} \\[2mm]
	\godTrans{(A \Lolly B)} & \equiv & (\godTrans{A} \iAnd (\godTrans{B})\Lnot)\Lnot.
\end{array}
\]
\end{Definition}
Recalling that $A\Lnot = A \Lolly 1$, one may check that $\kTrans{(A\Lnot)}$ and $\godTrans{(A\Lnot)}$ are equivalent to $(\kTrans{A})\Lnot$ and $(\godTrans{A})\Lnot$
respectively.

\begin{Theorem} \label{thm-kolm} Both the Kolmogorov translation $\kTrans{(\cdot)}$ and the G\"odel translation $\godTrans{(\cdot)}$ are double negation translations for $\ALi$.
\end{Theorem}
\Proof We sketch the proof for the Kolmogorov translation, leaving the G\"odel translation as an exercise for the reader. Clearly $\ALi + \DNE$ proves $A \simeq \kTrans{A}$, hence we have ({\bf DNS1}). Using the fact that $A\Lnot\Lnot\Lnot \simeq A\Lnot$ one can show $(\kTrans{A})\Lnot\Lnot \simeq \kTrans{A}$, and hence ({\bf DNS3}). Finally, in order to show ({\bf DNS2}) we prove a slightly stronger result, that if $\Gamma \vdash A$ is provable in $\ALc$ then $ \vdash \kTrans{(\Gamma \Lolly A)}$ is provable in $\ALi$, where $\Gamma \Lolly A$ abbreviates $B_0 \Lolly \ldots \Lolly B_n \Lolly A$. This can easily be shown by induction on the derivation of the sequent $\Gamma \vdash A$, noting that $(A\Lnot\Lnot \Lolly B\Lnot\Lnot)\Lnot\Lnot \simeq A\Lnot\Lnot \Lolly B\Lnot\Lnot$ and $A\Lnot\Lnot\Lnot \simeq A\Lnot$ are both provable in $\ALi$. \Done

\Subsection{Gentzen and Glivenko translations of $\LLc$ into $\LLi$}

For both the Gentzen and the Glivenko translations (defined below) a corresponding Theorem \ref{thm-kolm} no longer holds. These translation rely on uses of contraction which are not available in affine logic. Nevertheless, we show that the amount of contraction available in {\L}ukasiewicz logic is sufficient for these translations to go through.

\newcommand{\ggTrans}[1]{{#1}^{\sf Gen}}

\begin{Definition}[Gentzen translation \cite{Gentzen(33)}] \label{def:gentzen} Let ${\cal L}$ be the language of $\LLi$ and $\LLc$. For each formula $A \in {\cal L}$ associate a formula $\ggTrans{A} \in {\cal L}$ inductively as follows:
\[
\begin{array}{rcl}
	\ggTrans{P} & \equiv & P\Lnot\Lnot \quad\quad (\mbox{$P$ atomic}) \\[2mm]
	\ggTrans{1} & \equiv & 1 \\[2mm]
	\ggTrans{(A \iAnd B)} & \equiv & \ggTrans{A} \iAnd \ggTrans{B} \\[2mm]
	\ggTrans{(A \Lolly B)} & \equiv & \ggTrans{A} \Lolly \ggTrans{B}.
\end{array}
\]
Recall that $A\Lnot = A \Lolly 1$, hence we have that $\ggTrans{(A\Lnot)}$ is equivalent to $(\ggTrans{A})\Lnot$.
\end{Definition}

\begin{Theorem} \label{gentzen-soundness} The Gentzen translation $\ggTrans{(\cdot)}$ is a double negation translation for $\LLi$.
\end{Theorem}
\Proof We show by induction on the structure of $A$ that $\ggTrans{A}$ is equivalent over $\LLi$ to $\kTrans{A}$. The result will then follow by Theorem \ref{thm-kolm}. The non-trivial cases are implication and conjunction. For implication we have
\begin{align*}
\ggTrans{(A \Lolly B)} & \simeq \ggTrans{A} \Lolly \ggTrans{B} \by{def $\ggTrans{(\cdot)}$} \\[1mm]
	& \simeq \kTrans{A} \Lolly \kTrans{B} \by{IH} \\[1mm]
	& \simeq ((\kTrans{A})\Lnot\Lnot \Lolly (\kTrans{B})\Lnot\Lnot) \by{({\bf DNS3}) for $\kTrans{(\cdot)}$} \\[1mm]
	& \simeq (\kTrans{A} \Lolly \kTrans{B})\Lnot\Lnot \by{Theorem \ref{thm-lolly-not-not}} \\[1mm]
	& \simeq \kTrans{(A \Lolly B)}. \by{def $\kTrans{(\cdot)}$}
\end{align*}
Similarly for conjunction in this case using Theorem \ref{main-thm-n-trans}. \Done

\begin{Theorem}\label{thm:gentzen-not-ali} The  translation $\ggTrans{(\cdot)}$ is not a double negation translation for $\ALi$.
\end{Theorem}
\Proof We show that (DNS2) fails for the Gentzen translation on $\ALi$. Let $P, Q$ be atomic formulas. Note that the Gentzen translation of $(P \iAnd Q)\Lnot\Lnot \Lolly (P \iAnd Q)$ (an instance of $\DNE$) is
\[ (P\Lnot\Lnot \iAnd Q\Lnot\Lnot)\Lnot\Lnot \Lolly (P\Lnot\Lnot \iAnd Q\Lnot\Lnot) \]
but that is not provable in $\ALi$, as shown in \cite{arthan-oliva14b}.
\Done

\begin{Definition}[Glivenko translation \cite{Glivenko(29)}] \label{def:glivenko} Given a formula $A \in {\cal L}$ define its Glivenko translation $\glTrans{A}$ as $\glTrans{A} = A\Lnot\Lnot$.
\end{Definition}

\begin{Theorem} The Glivenko translation $\glTrans{(\cdot)}$ is a double negation translation for $\LLi$.
\end{Theorem}
\Proof Similar to the proof of Theorem \ref{gentzen-soundness}, by induction on the structure of $A$ we can show that $\kTrans{A} \simeq A\Lnot\Lnot$, i.e. $\kTrans{A} \simeq \glTrans{A}$. \Done

\begin{Theorem}\label{thm:glivenko-not-ali}
The Glivenko translation is not a double negation translation for $\ALi$.
\end{Theorem}
\Proof Let $P$ be an atomic formula. Note that the Glivenko translation of $P\Lnot\Lnot \Lolly P$ (an instance of $\DNE$) is $(P\Lnot\Lnot \Lolly P)\Lnot\Lnot$, which, as shown in \cite{arthan-oliva14b}, is not provable in $\ALi$.
Hence (DNS2) fails for the Glivenko translation for $\ALi$.
\Done

\begin{Theorem} There are extensions $\VA_1$ and $\VA_2$ of $\ALi$ such that
\begin{itemize}
	\item $\glTrans{(\cdot)}$ is a double negation translation for $\VA_1$ but $\ggTrans{(\cdot)}$ is not. 
	\item $\ggTrans{(\cdot)}$ is a double negation translation for $\VA_2$ but $\glTrans{(\cdot)}$ is not. 
\end{itemize}
\end{Theorem}
\Proof In \cite{arthan-oliva14b} we construct two finite models whose theories have the property above. \Done

\Section{Concluding Remarks}
\label{sec:conclusion}

We have presented {\L}ukasiewicz logic as an extension of minimal affine logic, and studied two important fragments: \emph{minimal {\L}ukasiewcz logic} and \emph{intuitionistic {\L}ukasiewcz logic}. We have shown that, quite surprisingly, several theorems of full intuitionistic (respectively, minimal) logic already hold in intuitionistic (respectively, minimal) {\L}ukasiewcz logic, where only a limited form of contraction $\CWC$ is available. Crucial tools in the study of these system have been the four derived connectives: weak conjunction ($\CCb{A}{B}$), strong disjunction ($\CKb{A}{B}$), strong implication ($\CMb{A}{B}$), and weak NOR ($\oCKb{A}{B}$). Several properties of the derived connectives have been established, including homomorphism properties for double negation, and De Morgan dualities. 

We believe that much stronger properties about the derived connectives hold in $\LLm$ and $\LLi$, and that avoiding the use of $\EFQ$ (and hence staying in the minimal setting) one might be able to generalise several of the results proven here. For instance, we have been able to show model-theoretically (cf. our companion paper \cite{arthan-oliva14b}) that Theorems \ref{thm-c-demorgan} and \ref{theorem-k-contradiction} hold in greatter generality. For Theorem \ref{thm-c-demorgan} one can show that the full (un)currying between the strong implication $\CM$ and the weak conjunction $\CC$ hold in minimal {\L}ukasiewcz logic, i.e.
\[ \CMb{(\CCb{A}{B})}{C} \simeq \CMb{A}{(\CMb{B}{C})} \]
Theorem \ref{thm-c-demorgan} is the case when $C = 1$. But we have no proof-theoretic proof of this as yet. Similarly, a more general version of Theorem \ref{theorem-k-contradiction}, that a $k$-contradiction implies double negation elimination, can be shown to hold (model-theorectically) in $\LLm$, namely,
\[ (A^n \Lolly B) \iAnd (\CKb{B}{A}) \vdash (\CKb{A}{B}) \]
for all $n \geq 1$. Theorem \ref{theorem-k-contradiction} is also the special case when $B = 1$, and $\EFQ$ is used to simplify $\CKb{A}{1} \simeq A$. But again, we have at present no proof-theoretic derivation of this result.

Let us conclude by noting that the consequence relation in the logic of commutative GBL-algebras has been recently shown to be PSPACE-complete \cite{Bova:2009}. Given that hoops are sub-reducts of such algebras\footnote{We thank Franco Montagna to pointed out these results to us.}, it follows by a result of Blok-Ferreirim that the complexity of deciding the consequence relation in the logic of (bounded) hoops (i.e. $\LLm$ and $\LLi$) is also in PSPACE. Nevertheless, we have found that the heuristics employed by Prover9 and Mace4, and by our algebraic methods (cf. \cite{arthan-oliva14b}), has worked surprisingly well in deciding validity of reasonably complex formulas. For instance, though with no a-priori bound on the search time, it turned that all proofs we found using Prover9 were discovered within 120 minutes of starting the program, while Mace4 explores all hoops of size 30 within a few minutes. 

\bibliographystyle{plain}

\bibliography{references}

%

\end{document}